\documentclass[aps,prl,reprint,amssymb,showpacs,superscriptaddress,notitlepage,onecolumn]{revtex4-2}
\usepackage{bm}
\usepackage{graphicx}
\usepackage{hyperref}
\usepackage{dcolumn}
\usepackage{braket}
\usepackage{natbib}
\usepackage{amsmath}
\usepackage{color}
\usepackage{mathtools,amssymb}
\usepackage{soul}
\usepackage{float}
\usepackage{amssymb}
\usepackage{esint}
\usepackage{mathtools}
\usepackage{physics}
\usepackage{makecell}
\usepackage{multirow}
\usepackage{array}
\usepackage{geometry}
\usepackage{url}
\usepackage{adjustbox}

\begin{document}
\title{Multi-timescale Frequency–Phase Matching for High-Yield Nonlinear Photonics}



\author{Mahmoud Jalali Mehrabad†}\email{mjalalim@umd.edu}
\affiliation{Joint Quantum Institute, University of Maryland and National Institute of Standards and Technology,
College Park, MD 20742, USA}

\author{Lida Xu}
\altaffiliation{Equally contributing authors}
\affiliation{Joint Quantum Institute, University of Maryland and National Institute of Standards and Technology,
College Park, MD 20742, USA}

\author{Gregory Moille}
\affiliation{Joint Quantum Institute, University of Maryland and National Institute of Standards and Technology,
College Park, MD 20742, USA}

\author{Christopher J. Flower}
\affiliation{Joint Quantum Institute, University of Maryland and National Institute of Standards and Technology,
College Park, MD 20742, USA}

\author{Supratik Sarkar}
\affiliation{Joint Quantum Institute, University of Maryland and National Institute of Standards and Technology,
College Park, MD 20742, USA}

\author{Apurva Padhye}
\affiliation{Joint Quantum Institute, University of Maryland and National Institute of Standards and Technology,
College Park, MD 20742, USA}

\author{Shao-Chien Ou}
\affiliation{Joint Quantum Institute, University of Maryland and National Institute of Standards and Technology,
College Park, MD 20742, USA}

\author{Daniel G. Suarez-Forero}
\affiliation{Joint Quantum Institute, University of Maryland and National Institute of Standards and Technology,
College Park, MD 20742, USA}

\author{Mahdi Ghafariasl}
\affiliation{Joint Quantum Institute, University of Maryland and National Institute of Standards and Technology,
College Park, MD 20742, USA}

\author{Yanne Chembo}
\affiliation{
Institute for Research in Electronics and Applied Physics, University of Maryland, College Park, MD 20742, USA}

\author{Kartik Srinivasan}
\affiliation{Joint Quantum Institute, University of Maryland and National Institute of Standards and Technology,
College Park, MD 20742, USA}

\author{Mohammad Hafezi}\email{hafezi@umd.edu}
\affiliation{Joint Quantum Institute, University of Maryland and National Institute of Standards and Technology,
College Park, MD 20742, USA}

\begin{abstract}


Integrated nonlinear photonic technologies, even with state-of-the-art fabrication with only a few nanometer geometry variation, face significant challenges in achieving wafer-scale yield of functional devices~\cite{wang2024lithium,zhang2025ultrabroadband,aghaee2025scaling,AlexanderNature2025,yanagimoto2025programmable,xin2025wavelength,dutt2024nonlinear}. A core limitation lies in the fundamental constraints of energy and momentum conservation laws. Imposed by these laws, nonlinear processes are subject to stringent frequency and phase matching (FPM) conditions that cannot be satisfied across a full wafer without requiring a combination of precise device design and active tuning.
Motivated by recent theoretical~\cite{mittal2021topological,huang2024hyperbolic,hashemi2024floquet,tusnin2023nonlinear,hashemi2025reconfigurable} and experimental~\cite{pang2025versatile,flower2024observation,xu2025chip} advances in integrated multi-timescale nonlinear systems, we revisit this long-standing limitation and introduce a fundamentally relaxed and passive framework: nested frequency-phase matching. As a prototypical implementation, we investigate on-chip multi-harmonic generation in a two-timescale lattice of commercially available silicon nitride (SiN) coupled ring resonators, which we directly compare with conventional single-timescale counterparts. We observe distinct and striking spatial and spectral signatures of nesting-enabled relaxation of FPM. Specifically, for the first time, we observe simultaneous fundamental, second, third, and fourth harmonic generation, remarkable 100~\% multi-functional device yield across the wafer, and ultra-broad harmonic bandwidths. Crucially, these advances are achieved without constrained geometries or active tuning, establishing a scalable foundation for nonlinear optics with broad implications for integrated frequency conversion and synchronization, self-referencing, metrology, squeezed light, and nonlinear optical computing.
\end{abstract}

\maketitle

\section{Introduction}

The conservation laws of energy and momentum impose stringent constraints on parametric nonlinear optical processes, where simultaneous frequency and phase matching (FPM) is essential for efficient frequency conversion~\cite{boyd2008nonlinear}. A prominent example is harmonic generation (HG), wherein light-matter interaction produces new frequencies at integer multiples of the pump~\cite{winterfeldt2008colloquium}. In addition, parametric nonlinear processes can create multiple entangled photons from single pump photons, an essential ingredient for quantum networks~\cite {KimbleNature2008}. The advent of nonlinear integrated photonics in technologically mature materials, such as silicon nitride (SiN)~\cite{HeckLaserPhotonicsRev.2014,liu2021high}, that support low-loss, wafer-scale foundry fabrication should enable such classical and quantum technologies to expand outside of the lab and toward commercial applications. In the context of HG in SiN, major milestones include the use of intrinsic and effective $\chi^{(2)}$ nonlinearities to enable second-harmonic generation(SHG)~\cite{levy2011harmonic,billat2017large,porcel2017photo,lu2021efficient,nitiss2022optically} and spontaneous parametric down-conversion~\cite{DalidetOpt.ExpressOE2022, li2025down}. However, while fabrication process maturity has begun to reach a level where geometric control at the few-nanometer level across the wafer is feasible, this has not been translated to wafer-scale functional device yield, presenting a significant bottleneck for commercial viability~\cite{aghaee2025scaling}.
Specifically, the central challenge is the strict FPM conditions for nonlinear optical processes, where precise compensation of the intrinsic material dispersion through waveguide geometry engineering across large bandwidths is required. Those strict FPM conditions can make even a few nanometer dimensional variations result in poor yield unless active tuning strategies are employed, particularly in the context of resonant geometries that are often used to enhance efficiency.

This problem is related to the fact that integrated cavities such as microring resonators typically operate under a single timescale set by the cavity round-trip, resulting in discretized frequencies and momenta (\textit{i.e.,} azimuthal mode numbers)~\cite{dutt2024nonlinear,lu2021efficient,li2023high}. 
The single timescale characteristic makes FPM in resonators even more sensitive to geometrical dimensions than in waveguides~\cite{lu2021efficient}, preventing high functional device yields even in mature foundry platforms like SiN, thus hindering commercial scale-up. Post-fabrication FPM tuning strategies such as integrated heaters~\cite{li2023high} can provide a solution, but their large static power consumption and thermal cross-talk can limit device scaling and complexity. Moreover, all-optical poling can yield a substantial  $\chi^{(2)}_{\mathrm{eff}}$ in SiN which relaxes phase-matching considerations~\cite{nitiss2022optically}, but frequency matching is still needed, and this mechanism does not extend to $\chi^{(3)}$ processes. Thus, a strategy that intrinsically and passively relaxes FPM constraints has remained elusive. Recently, nested resonator architectures with independently tunable timescales have emerged, enabling coarse–fine frequency grids~\cite{mittal2021topological,huang2024hyperbolic,hashemi2024floquet,tusnin2023nonlinear,hashemi2025reconfigurable}. These systems have demonstrated new capabilities in frequency comb generation~\cite{flower2024observation}, mode-locking~\cite{xu2025chip}, and enhanced entanglement generation~\cite{pang2025versatile}, opening the door to revisiting FPM in a multi-timescale framework.

In this work, we introduce and experimentally demonstrate a two-timescale (nested) FPM mechanism, which we prototypically leverage for passive on-chip multi-harmonic generation with exceptional bandwidth, tunability, and fabrication tolerance. Using nested arrays of coupled SiN ring resonators, fabricated in a commercial photonic foundry, with independently controllable fast and slow timescales, we generate a third harmonic via $\chi^{(3)}$ nonlinearities at telecom and visible wavelengths, while simultaneously realizing broadband second harmonic in the near-visible via a photo-induced $\chi^{(2)}_{\mathrm{eff}}$. The resulting spatial and spectral signatures reveal harmonic generation spanning two octaves over 1~{\textmu}m (from telecom to visible wavelengths), robust operation across chips with varying disorder landscapes, and versatile tunability via the nested two-timescale degree of freedom—all without post-fabrication tuning or engineered mode alignment. Moreover, simultaneous with the second and third harmonics, we observe the generation of light in the fourth harmonic band in our SiN devices, which is the first observation of such a process on an integrated SiN platform, serving as another signature of relaxed FPM. These results establish nested FPM as a new paradigm for passive and robust nonlinear photonic phenomena, with wide-ranging applications in both classical and quantum regimes, such as nonlinear optical computing.

\begin{figure*}[t]
\centering
\includegraphics[width=0.99\textwidth]{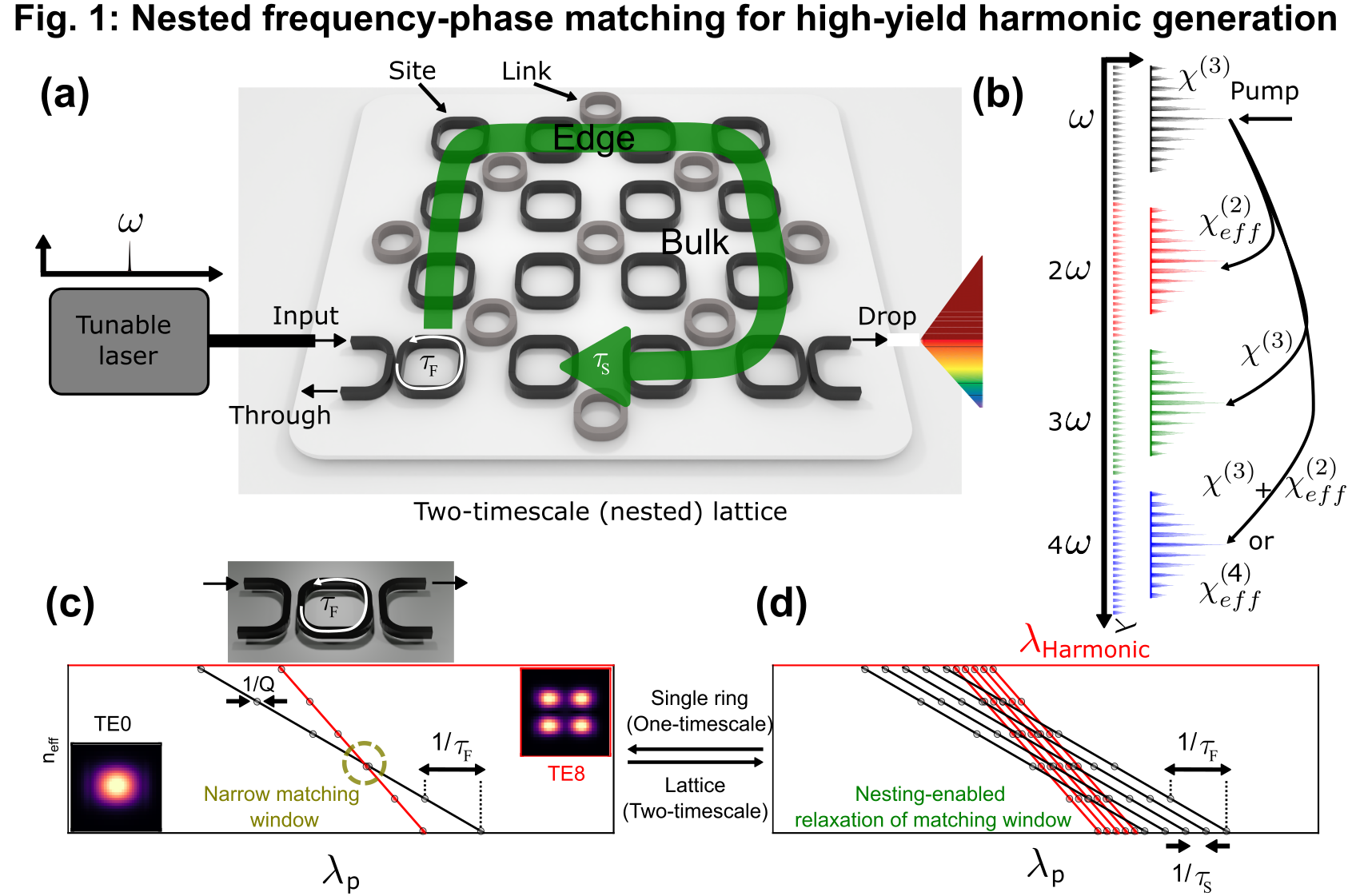}
\caption{\textbf{(a)} Schematic of nested harmonic generation in a simplified 2D array of coupled SiN resonators (see Figure S1 for actual device). A tunable telecom pump is coupled into the lattice at the input port and circulates along the edge of the 2D AQH SiN lattice (clockwise mode shown). Generated harmonics are analyzed via optical spectrum analyzers (OSAs), revealing the two characteristic timescales: $\tau_{\rm{F}}$ of the individual rings and $\tau_{\rm{S}}$ of the super-ring. Harmonics are also imaged with infrared and visible cameras (not shown). \textbf{(b)} Schematic of the nonlinear processes and wavelength regimes for each harmonic band (see Figure S1 for details of all the other possible nonlinear pathways). Phase-matching illustration for SHG in \textbf{(c)} a single-timescale ring vs. \textbf{(d)} two-timescale lattice (right). Insets: simulated fundamental and higher order TE mode profiles of an 800 nm \texttimes{} 1200 nm SiN waveguide at fundamental and SH wavelengths (see SI section S2 for detailed FDTD mode simulations). In the nested case (right), discrete single-ring phase-matching points are replaced with a two-timescale grid set by $\tau_{\rm{S}}$ (note that the two timescales are not to scale). Circle size reflects the mode linewidth inversely proportional to its quality factor Q.}
\label{Fig:intro}
\end{figure*}

\section{Concept and Experimental Scheme}

Figure~\ref{Fig:intro} presents the central concept of two-timescale FPM using simultaneous multi-harmonic generation in SiN ring resonators as an illustrative example. The device comprises a 2D array of coupled SiN ring resonators designed to emulate the anomalous quantum Hall (AQH) model for photons~\cite{leykam2018reconfigurable,mittal2019photonic,mittal2021topological,jalali2023topological}. Below, we highlight the key features relevant to the nested FPM mechanism; comprehensive device design details are provided in the methods section M1.

\subsection{Two-Timescale System and Harmonic Generation Scheme}

A defining feature of this platform is its nested frequency structure, enabled by two independently tunable timescales. The fast timescale $\tau_{\rm{F}}$ is set by the single-ring roundtrip time and corresponds to the mode spacing of individual rings $\Omega_{\rm F} = 1/\tau_{\rm F}$ (black), while the slower $\tau_{\rm{S}}$ arises from edge state spacing in the super-ring lattice $\Omega_{\rm S} = 1/\tau_{\rm S}$ (green), as shown in Figure~\ref{Fig:intro}a. The slower timescale $\tau_{\rm{S}}$ is tunable via inter-ring coupling strength~\cite{mittal2021topological,xu2025chip}, and its mode count is determined by the array size (see methods section M2).

Harmonic generation is driven by two nonlinear mechanisms in SiN waveguides. The Kerr ($\chi^{(3)}$) nonlinearity facilitates the four-wave mixing process (near the pump at telecom wavelength) for the fundamental and (in the visible wavelength range) third harmonic generation (THG). Simultaneously, an all-optical poling process, enabled by the coherent photogalvanic effect~\cite{levy2011harmonic,billat2017large,porcel2017photo,lu2021efficient,nitiss2022optically}, yields effective $\chi^{(2)}_{\mathrm{eff}}$ for second harmonic generation (and sum-frequency generation). Moreover, we observe light generation in the fourth harmonic band (Fig.~\ref{Fig:intro}b), likely due to cascaded $\chi^{(2)}_{\mathrm{eff}}$ and/or $\chi^{(2)}_{\mathrm{eff}}$+$\chi^{(3)}$ processes, though a $\chi^{(4)}_{\mathrm{eff}}$ nonlinearity is also in principle possible. See Figure S1 for all the possible nonlinear pathways.

The generation and detection are enabled by pumping the system at resonance via a tunable telecom laser, with harmonics collected from the drop port and analyzed using optical spectrum analyzers. Additional spatial imaging of the harmonics is performed using spectrally filtered visible and IR cameras (not shown here; see Supplementary Information (SI) section S1).

\subsection{Single-Timescale vs. Nested FPM}

Figure~\ref{Fig:intro}c,d contrasts the conventional single-timescale with our nested approach using the SHG phase-matching as an example (here assuming perfect phase-matching without the benefit of all-optical poling). Each panel plots the group index vs. pump (bottom x-axis) and SH wavelength (top x-axis), with discrete ring resonances at both wavelengths shown as black and red circles. Insets show simulated fundamental TE$_0$ and TE$_8$ mode profiles for SiN waveguides at corresponding wavelengths. Details of the geometry and simulations are in the SI section S2.

In single-timescale rings, FPM is restricted to narrow spectral overlaps, shown as a dashed circle in Figure~\ref{Fig:intro}c. Deviations from modal alignment quickly degrade SHG efficiency. We note that achieving such alignment becomes increasingly harder the higher the harmonic is, and for modes with larger quality factors (narrower spectral linewidth).

In contrast, nested lattices (Figure~\ref{Fig:intro}d) feature a denser, independently tunable $\tau_{\rm{S}}$ mode grid (both the mode count and mode spacing are readily controllable by design, see methods section M2). For our experiments, $\tau_{\rm{F}} \approx$ 1~THz and $\tau_{\rm{S}} \approx$ 3~GHz, yielding a two-dimensional grid of FPM points. This dramatically expands the phase-matching window compared to the isolated resonance condition of single rings. Additionally, $\tau_{\rm{S}}$ spacing and mode count provide a new degree of freedom for tailoring the FPM landscape.

In the following sections, we experimentally investigate nested FPM through on-chip multi-harmonic generation and analyze its consequences for device yield, bandwidth, and tunability.

\section{Results and Analysis}

\subsection{Measurement Setup}

Our two-timescale device is a 10 \texttimes{} 10 array of coupled SiN ring resonators in an add-drop configuration. Each waveguide has a cross-section of 800 nm \texttimes{} 1200 nm. The coupling gaps between the input-output waveguides and the resonators are designed to yield a coupling rate of $2\pi\times30$~GHz at the pump mode and wavelength. We note that our harmonic generation process leverages FPM between the fundamental mode at the pump and higher order transverse modes at harmonic wavelengths (see simulated mode profiles of our waveguide in the SI section S2), which enables efficient extraction of light at the frequency of interest for each harmonic despite a 2 to 3 octave working bandwidth. Moreover, the designed values for the two-timescales $\tau_{\rm{F}}$ and $\tau_{\rm{S}}$ result in free spectral ranges (FSRs) of 750~GHz and 2.5~GHz, corresponding to 6.3~nm and 20~pm, respectively, at telecommunication wavelengths. These correspond to longitudinal modes indexed by $\mu$ and $\sigma$, respectively. The lattice supports approximately 20 $\sigma$-modes across a 400~pm  (50~GHz) edge bandwidth, which repeats every 6.3~nm (single-ring FSR) interval. Detailed simulated and measured linear characterization is provided in the SI section S3.

For the generation of the harmonics in our two-timescale device, we operate in the quasi-continuous-wave excitation regime, which has been successfully employed in prior demonstrations of frequency comb generation in large coupled-ring arrays~\cite{flower2024observation,xu2025chip}. For spectral and spatial detection of the harmonics, we use broadband OSAs and cameras covering from the visible to the telecommunication bands. Experimental setup details are available in the SI section S1.

\subsection{Spatial and spectral signatures of two-timescale harmonics}

\begin{figure*}[t]
    \centering
    \includegraphics[width=0.83\textwidth]{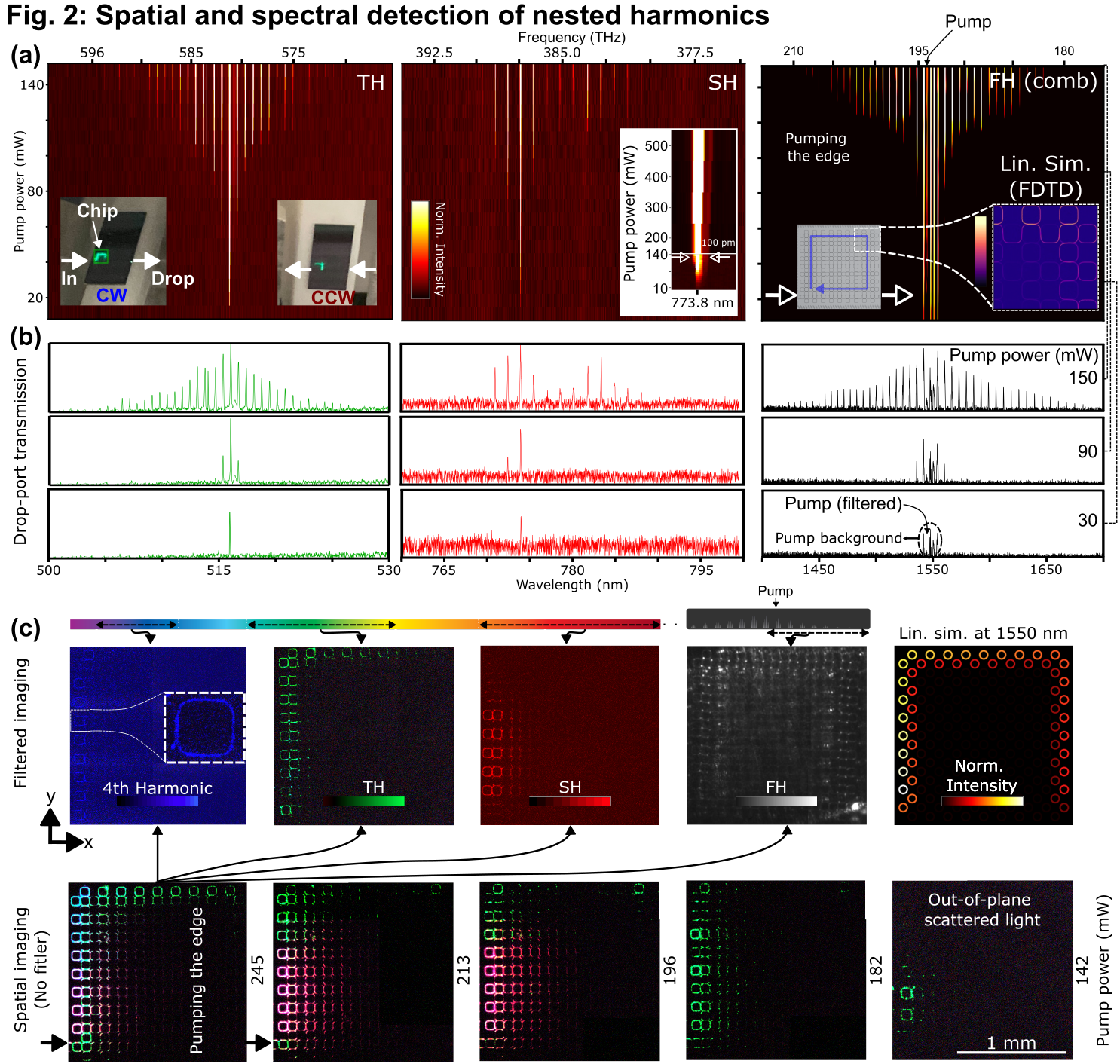}
    \caption{\textbf{(a)} From right to left: Measured pump-power-dependent spectra of the fundamental, second, and third harmonic (FH, SH, and TH, respectively) bands from the drop port of the 10 × 10 SiN lattice, when pumping the center of the edge band. The inset in the SH panel shows a zoom-in of the pump-power dependence of the SH signal near 776.3 nm, with the 100 pm bandwidth indicated by the white arrows. An optical image of the device, indicating the input and drop ports, as well as an FDTD simulation of part of a typical edge profile, is shown in the inset of the FH panel (the color bar indicates the normalized electric field intensity). Insets in the TH panel show visible green emission from the device edges in both clockwise (CW) and counter-clockwise (CCW) directions. \textbf{(b)} Examples of the corresponding FH, SH, and TH spectra at different pump powers. \textbf{(c)} Bottom row: unfiltered real-color pump-power-dependent spatial imaging of the harmonics, when pumping the edge at 1548.60~nm. Top row, from right to left: simulated linear intensity profile of a typical edge mode, followed by real-color spectrally filtered FH, SH, TH, and the fourth harmonics, respectively.}
    \label{Fig:imaging}
\end{figure*}

We initiate the generation of nested harmonics by pumping the array at the center of one of its edge bands, centered at 1548~nm, hereafter referred to as longitudinal mode $\mu = 0$. The generated harmonics are collected at the drop port of the device as a function of pump power and spectrally analyzed using two broadband optical spectrum analyzers (OSAs), collectively covering the 400~nm to 1800~nm range. The results are presented in Figure~\ref{Fig:imaging}a. As the pump power is increased, near the pump wavelength (right panel), we sequentially observe the generation of sidebands through spontaneous four-wave mixing (FWM) followed by frequency comb formation via cascaded FWM at an optical parametric oscillation (OPO) threshold of approximately 100~mW average pump power. The comb profile is preserved and broadened at higher pump powers. For noise analysis of the combs see the SI section S4.

Simultaneously, we observe the generation of SH and TH (Figure~\ref{Fig:imaging}a, the middle and left panels, respectively). At the lowest pump power, before the onset of FWM in the pump band, we observe single harmonic tones in the SH and TH bands. Subsequently, we study the evolution of the spectra across the SH and TH bands with pump power and observe similar behavior, though with a few harmonic-order-dependent distinct characteristics. Notably, the fact that multi-harmonic generation closely follows that of the FH and its OPO onset provides initial experimental evidence for relaxed nested FPM conditions in the two-timescale lattice. Moreover, as shown in the inset of Figure~\ref{Fig:imaging}a (SH plot, middle panel), upon increasing pump power, we observe the population of the entire 100~pm (50~GHz) bandwidth of the edge band near the SH band (same as the 50~GHz edge bandwidth at the pump wavelength), serving as the first signature of nesting characteristics of the harmonics. We investigate this spectral characteristic of nesting in detail in the next section. Moreover, we note that while the TH spectra closely follow that of the FH, the SH spectrum exhibits a more complex structure, featuring multiple local maxima and broadening asymmetrically more towards the longer wavelength side at higher pump powers. This behavior likely arises from differences between the optically induced $\chi^{(2)}_{\mathrm{eff}}$ process responsible for SHG and the intrinsic $\chi^{(3)}$ nonlinearity governing FWM in the FH band and TH generation in SiN. A detailed optical power analysis of the harmonics is available in the SI section S5. To take a closer look at the spectral characteristics of the generated harmonics, we plot the harmonic spectra at a fixed pump power above its OPO threshold using 185~mW of average power, shown in Figure~\ref{Fig:imaging}b. The mode spacing between adjacent FH comb teeth is approximately 6.3 nm, corresponding to the single-ring free spectral range, $\Omega_{R} = 1/\tau_{\rm{F}}$. The spectral spacing of SH resonances is approximately 1.5 nm (compared to 3 nm expected for the pure SHG process). This observation, as well as the 0.67 nm mode spacing for TH spectra, is likely from additional nonlinear processes such as the sum-frequency generation (SFG) process, which in the SH band is due to mixing between the different FWM tones in the fundamental band, and in the TH band, is due to mixing between tones from the fundamental and SH bands. The inset of the TH spectrum (left panel) shows images of the chip, excited in both counterclockwise (CCW) and clockwise (CW) directions, where the edge-confined generated green light is clearly visible.

Next, we employ spectrally-resolved spatial imaging to demonstrate that the generated harmonics inherit the two-timescale characteristics of the linear system and are confined to the boundary of the AQH lattice. This direct imaging of edge harmonics is performed by capturing out-of-plane scattered light resulting from fabrication-induced imperfections and disorder. Scattered light is collected from above using an objective lens and imaged independently on infrared (IR) and visible cameras. For FH imaging, the pump signal (and part of the comb) is removed using spectral filters (1600~nm long-pass), and only FH emission is recorded. Moreover, we use the following filtering settings: 650~nm to 800~nm for SH, 500~nm to 600~nm for TH, and 450~nm short-pass filter for the fourth harmonic. The results are presented in Figure~\ref{Fig:imaging}c. First, before using the filters, (bottom row, from right to left) as a function of the pump power, we observe increasingly stronger and multi-color scattered light from the boundary of the array. Next, images (using a fixed pump power of 245~mW) are shown in Figure~\ref{Fig:imaging}c (top row). From right to left, the simulated linear intensity profile of an edge mode, followed by filtered harmonic-resolved real-color (aside from the FH light shown in gray color) spatial imaging results, are shown. We note that due to the strong coupling between the lattice and bus waveguides, the circulating light weakly persists beyond the first output port it encounters. It is also crucial to note that generally, our imaging of the scattered out-of-plane light should not be confused with drop-port transmission measurements (for imaging, simply by pumping the device with moderately higher power, scattered light can be observed all the way from the input to the drop port, as shown in Figure S5). Importantly, the edge propagation remains robust even around sharp 90$^\circ$ corners, with no appreciable scattering into the bulk. These observations confirm that the harmonics originate from within the edge band and that the topological protection persists under such a diverse family of nonlinear processes. Detailed pump-wavelength-dependent spatial imaging of the multi-harmonics is presented in the SI sections S6-S8.

Moreover, we emphasize that, for the first time, we observe the generation of light in the fourth harmonic band simultaneously with SH and TH in a SiN ring system (blue panel in Figure~\ref{Fig:imaging}c. Note that we only perform spatial imaging of the fourth harmonic since unlike our OSAs, our visible camera remains efficient at blue color). This striking and unprecedented observation reflects complex dynamics and potential multi-process nonlinear interactions in our system that may enable an effective fourth-order nonlinearity in our SiN lattice, but are yet to be fully understood. For example, light in the fourth harmonic band can be created through sum-frequency generation between two tones in the second-harmonic bands, sum-frequency generation between the fundamental and third-harmonic bands, or directly through a fourth-order nonlinearity. Regardless of the nonlinear pathway, our observation serves as yet another evidence of relaxed FPM in the system.

\begin{figure*}[t]
    \centering
    \includegraphics[width=0.9\textwidth]{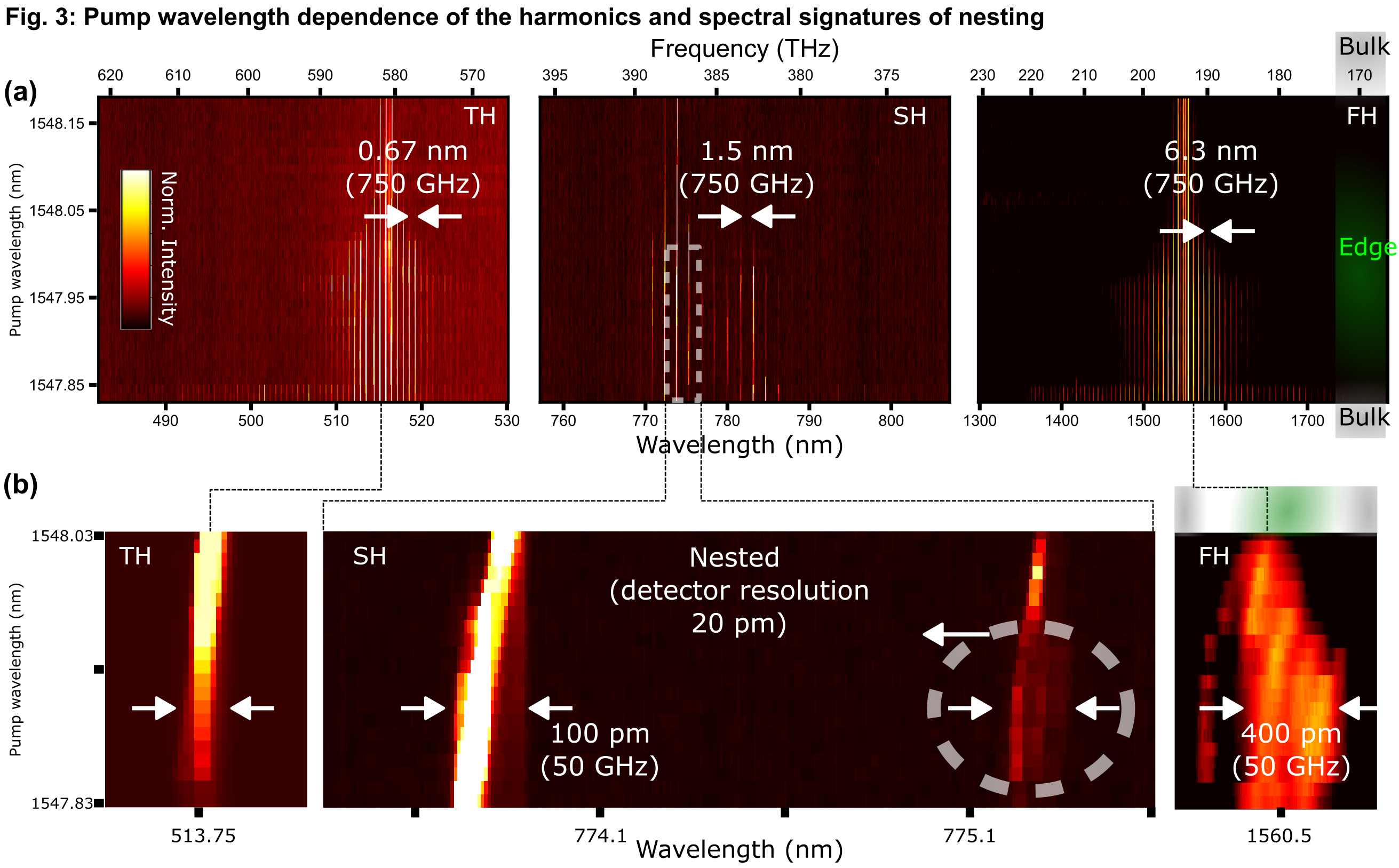}
    \caption{\textbf{(a)} From right to left: Broadband optical spectrum analysis of the nested fundamental, second, and third harmonics (FH, SH, and TH) as a function of pump wavelength, at approximately 185~mW average pump power. The bulk and edge regions of the spectrum are highlighted in gray and green, respectively. \textbf{(b)} Spectral analysis and bandwidth comparison of a representative FH (right), two adjacent SH modes (middle), and a typical TH fast-timescale resonance (left). The observed bandwidths—approximately 400~pm and 100~pm for FH and SH, respectively—indicate full occupation of the edge band at each longitudinal mode. The dashed circle in the SH panel highlights the nested substructure of the harmonics, limited by the 20~pm  resolution of the grating-based OSAs (see the SI section S9 for the high-resolution heterodyne-based spectral analysis of the nested FH).}
    \label{Fig:osa}
\end{figure*}

\subsection{Spectral Signatures of Two-Timescale Harmonics}

Next, to probe the spectral characteristics of nested harmonic generation, we tune the pump wavelength within the edge band over 350 pm to cover multiple slow timescale modes ($\sigma$'s).  A detailed linear characterization of the device is provided in SI Section S3. The resulting harmonic spectra, acquired using OSAs, are shown in Figure~\ref{Fig:osa}a. We find that pumping within the edge band produces harmonics with well-defined and comparable spectral envelopes. Notably, the observation of higher harmonics (SH and TH) with bandwidths corresponding to that of the FH provides additional evidence for the relaxation of FPM constraints via the nested two-timescale mechanism. A detailed analysis of the pump power and wavelength dependence of the harmonic generation is provided in the SI section S5.

To provide additional direct evidence of the nested structure of the harmonics within each single-ring resonance, we perform a detailed inspection of selected harmonic teeth, as shown in Figure~\ref{Fig:osa}b. Within each of these longitudinal modes, we observe full occupation of the edge band, with bandwidths scaling according to the harmonic order: approximately 400~pm for FH and 100~pm for SH (accurately estimating the bandwidth of TH is not feasible due to the limited resolution of the detector). 

Although the resolution of the grating-based OSAs is limited to approximately 20~pm, a clear substructure is visible within the harmonic spectra, indicative of nested frequency components. These features are consistent with prior observations in nested frequency comb systems~\cite{flower2024observation,xu2025chip}. In those studies, using comparable devices, high-resolution heterodyne-based spectroscopy with sub-picometer precision confirmed the presence of $\approx$20 nested resonances spanning the full edge band, in agreement with theoretical expectations. We confirm this for our FH spectra, see the SI section S9 for the high-resolution heterodyne-based spectral analysis of the nested FH.

\begin{figure*}[t]
    \centering
    \includegraphics[width=0.9\textwidth]{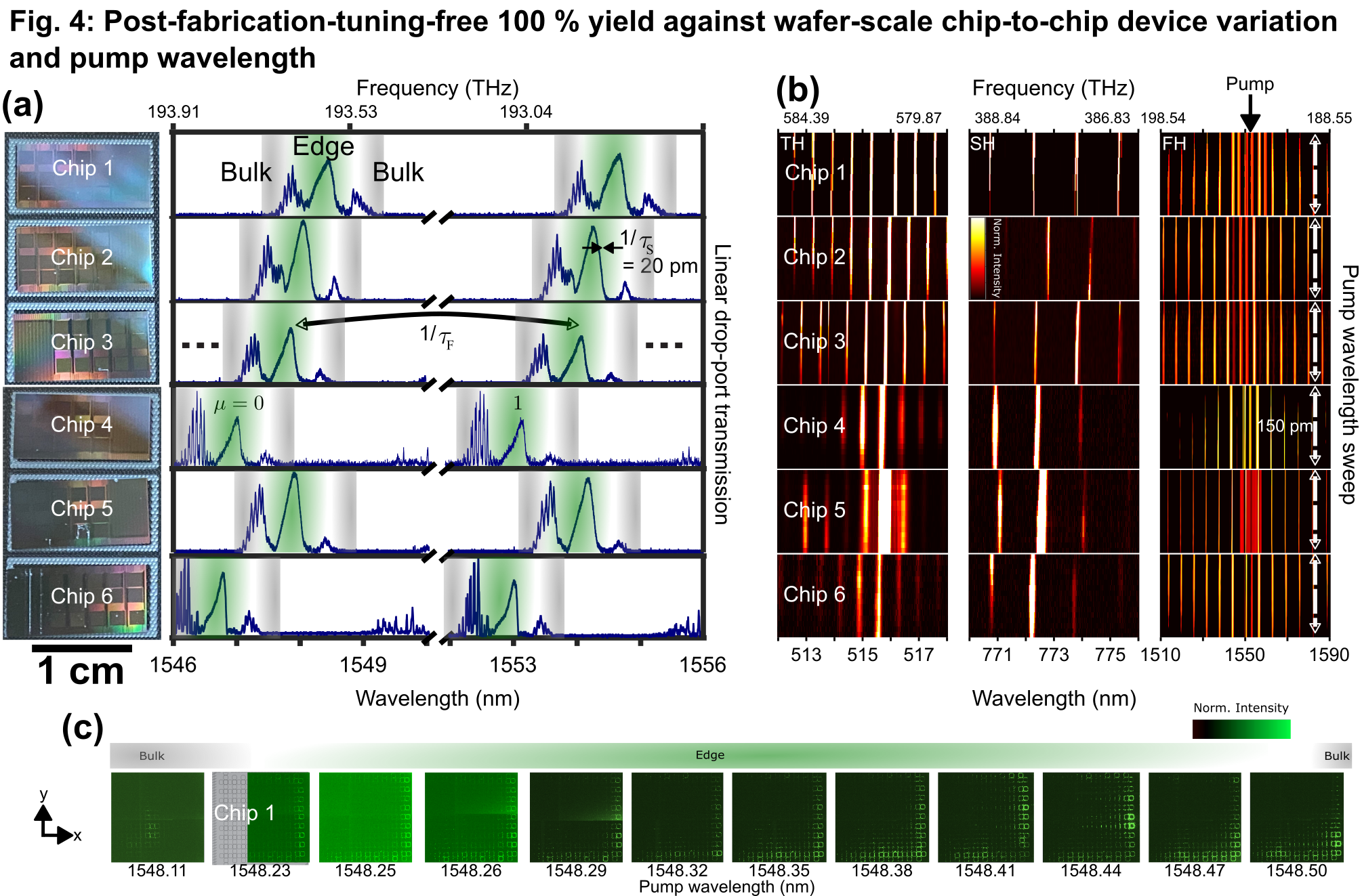}
    \caption{\textbf{(a)} Measured drop-port linear-regime spectra from six devices with identical designs fabricated on separate chips. The edge and bulk regions of the band are highlighted in green and gray, respectively. Significant chip-to-chip variations in spectral profiles and mode frequencies are observed. \textbf{(b)} Measured spectral evolution in the FH, SH, and TH bands for each chip, showing robust and comparable multi-harmonics simultaneously generated over a 150~pm wide pump wavelength sweep (y-axis) within the edge band of each device (to cover multiple slow timescale modes ($\sigma$'s) which are spaced by 20~pm), using approximately 185~mW of average pump power. \textbf{(c)} Pump wavelength-dependent true-color spatial imaging of TH light generated across the entire 400~pm-wide edge band, showing consistent edge confinement. In all TH images, a band-pass filter covering 500~nm to 600~nm was used. A bulk case at 1547.66 nm is shown for comparison in the leftmost image. The 1548.23~nm panel is partially overlayed on the optical image of the device for clarity.}
    \label{Fig:robust}
\end{figure*}

\subsection{Chip-to-Chip Fabrication Tolerance}

To further illustrate the robustness of nested FPM compared to conventional single-timescale approaches, we perform harmonic generation experiments on multiple devices with identical designs fabricated on separate chips through a multi-project wafer (MPW) fabrication. These experiments assess the tolerance of harmonic generation to typical fabrication-induced variation. The results are summarized in Figure~\ref{Fig:robust}.

Figure~\ref{Fig:robust}a shows the linear drop-port transmission spectra of a weak, broadband, tunable CW probe across six devices. As commonly seen even in state-of-the-art nanophotonic device fabrication and characterization, we observe notable chip-to-chip variation in spectral profiles, along with shifts in the relative longitudinal mode wavelength of up to $\approx$1.6~nm ($\approx$199~GHz). These significant and typical shifts are associated with differences in wafer thickness (an unavoidable few percent thickness variation in our 800 nm wafer case, for example) and disorder landscapes across chips.

Despite such variation, simultaneous multi-harmonic generation remains qualitatively consistent across all devices. The results for all six chips are shown in Figure~\ref{Fig:robust}b. For each device, we perform a 150~pm-wide pump wavelength sweep (y-axis) within the edge band and observe comparable multi-harmonics generated under similar pump powers. These results highlight the fabrication tolerance of nested FPM, in stark contrast to the narrow and sensitive phase-matching conditions typically encountered in single-ring systems (without thermal tuning). We further investigate the robustness of third-harmonic (TH) generation by performing spatial imaging in a separate device (Chip 1), distinct from the device (Chip 3) used in Figure~\ref{Fig:imaging}. Specifically, we conduct a detailed study of TH emission as a function of pump wavelength. The results are presented in Figures~\ref{Fig:robust}c. In all cases, the harmonic light was spectrally filtered using a 500~nm to 600~nm bandpass filter to isolate the TH signal.

\begin{figure*}[t]
    \centering
    \includegraphics[width=0.9\textwidth]{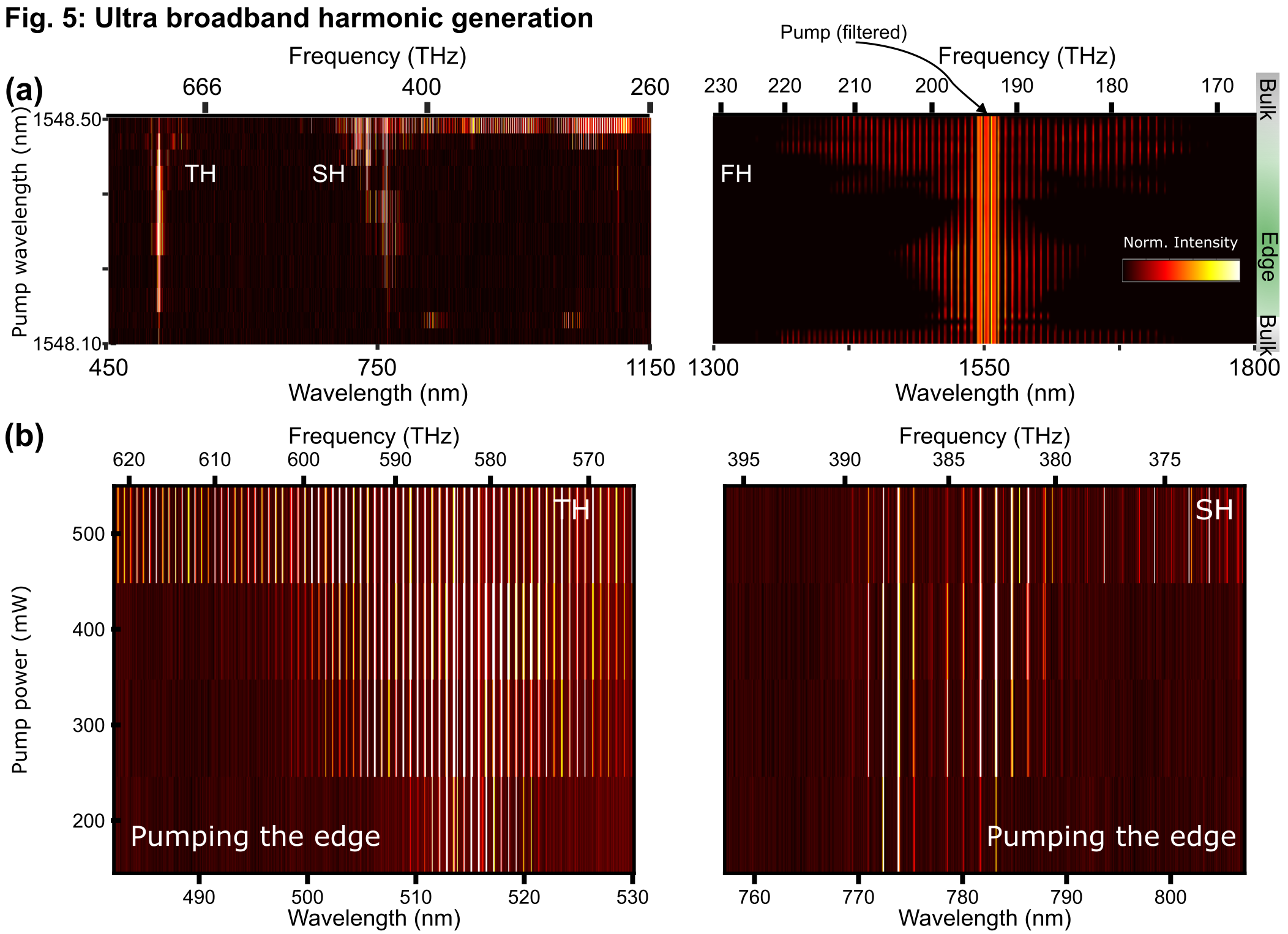}
    \caption{\textbf{(a)} Simultaneous ultra-broadband generation of (right) SH and (left) TH. The y-axis corresponds to a 100~pm pump wavelength sweep. The edge and bulk regions of the pump wavelength range are highlighted in green and gray, respectively. \textbf{(b)} Broadening the SH (right) and TH (left) spectra as a function of pump power when pumping the edge band.}
    \label{Fig:broad}
\end{figure*}

First, as shown in Figure~\ref{Fig:robust}c, pumping a bulk mode at 1548.11~nm—outside the edge band—results in TH emission localized exclusively within the bulk of the lattice, with no observable edge confinement. This provides a clear contrast between edge- and bulk-mode excitation. 

Moreover, we explore TH generation across the full 400~pm wide FH edge band, as summarized in Figure~\ref{Fig:robust}c. Strikingly, robust edge confinement of the green TH light is observed throughout the entire band, providing further evidence that the harmonic generation preserves the underlying topological protection of the lattice and remains resilient to device-level variations. Additional spatial imaging results across a wide range of pump parameters for both TH and FH are provided in the SI sections S3 and S7, respectively.

Next, we perform a direct comparative study of harmonic generation in the single-timescale single-ring counterparts of our system. We investigate both moderate (2000) and high (exceeding 1 million) quality factor devices, and as indicated in Figure~\ref{Fig:intro}c, observe very low device yield, narrow FPM windows, which were only possible using active tuning of the resonances, and lack of broadband operation. The results are presented in the SI sections S10-S11.

\subsection{Ultra broadband and tunable multi-harmonic generation}

Having established robust and high-yield nested multi-harmonic generation in our system, next, we explore several versatile and novel tuning mechanisms enabled by our relaxed FPM for controlling the spectral and spatial degrees of freedom of the harmonic. Figure~\ref{Fig:broad}a illustrates harmonic spectra generated within the FH and SH bands broadened beyond 450~nm by sweeping the pump from the edge band all the way close to adjacent bulk modes. Such unprecedentedly broad-bandwidth harmonic generation in a microresonator-based system reflects complex dynamics and potential multi-nonlinear processes that are yet to be understood.

Further bandwidth broadening is achieved by increasing pump power up to 500~mW, as shown in Figure~\ref{Fig:broad}b. Even moderate powers relative to the OPO threshold produce remarkable on-chip bandwidths, demonstrating the utility of nested FPM for integrated broadband light generation (compare this to the pump-power-dependent performance of single-ring, as shown in Figure S10).

Notably, all of these control mechanisms occur without any active post-fabrication device tuning, highlighting the intrinsic tunability and scalability of the platform.

\section{Summary and outlook}

In summary, we introduced a novel frequency and phase matching (FPM) mechanism based on multi-timescale frequency nesting, which inherently and passively relaxes the stringent constraints imposed by conventional single-timescale FPM on nonlinear optical processes. Our demonstration of ultra-broadband, two-timescale, highly tunable, and fabrication-tolerant multi-harmonic generation in a commercially available platform offers direct device yield advantages for general integrated nonlinear photonic technologies across many other platforms~\cite{lin2019broadband,wang2024lithium,zhang2025ultrabroadband,aghaee2025scaling,AlexanderNature2025,yanagimoto2025programmable,xin2025wavelength,dutt2024nonlinear,zang2025laser,moore2011continuous}. In particular, our approach can be highly attractive for higher refractive-index-contrast platforms, which generally exhibit even stricter degrees of FPM sensitivity than our SiN platform. Examples of specific near-term applications include on-chip multi-wave processes~\cite{li2025efficient,wang2022synthetic}, ultra-violet light generation~\cite{hwang2023tunable,moore2011continuous}, broadband comb self-referencing~\cite{moille2023kerr,Moille2024_arXivTRN} and spectroscopy~\cite{suh2016microresonator,coddington2016dual,herman2025squeezed,hariri2024entangled}, astrocombs~\cite{obrzud2019microphotonic,suh2019searching}, coarse-fine grid frequency metrology, and broadband spectroscopy. The nested harmonic structure is particularly suited for atomic clock interrogation and interfacing with narrow-linewidth transitions. Furthermore, emerging technologies such as Pockels lasers~\cite{bruch2021pockels} and parametrically driven soliton formation~\cite{englebert2021parametrically,moille2024parametrically} stand to benefit from this unique, multi-timescale FPM approach. 

Beyond this prototypical demonstration, our work introduces a paradigm shift in general FPM engineering, unlocking new opportunities across both classical and quantum nonlinear optics as well as nonlinear optical computing. In particular, nested FPM is in particular advantageous for very recent developments on integrated nonlinear optical computing~\cite{li2025all,basani2024all,bandyopadhyay2024single,yanagimoto2025programmable}, which are largely enabled by on-chip nonlinear optical processes. Moreover, beyond the classical regime, the relaxed FPM conditions enabled by nesting provide a promising route toward robust on-chip entangled photon sources~\cite{li2025down}, squeezed light~\cite{jahanbozorgi2023generation,shen2025strong,yang2021squeezed,shen2025highly}, quantum-correlated photon generation~\cite{zhao2023quantum}, exploring noise-free quantum frequency conversion protocols~\cite{lu2021proposal} and nonlinear gates~\cite{krastanov2021room}, and synchronized OPO networks for implementing nonlinear Ising machines~\cite{mcmahon2016fully,marandi2014network,honjo2021100,roy2022temporal}. Moreover, being free from post-fabrication-tuning strategies such as heaters, challenges of photonic-electric integration~\cite{atabaki2018integrating} are innately mitigated by nested FPM.

Finally, on a fundamental level, several intriguing steps forward stand out. One is gaining a deeper understanding of the interplay between nonlinearity and nested lattice physics~\cite{mittal2021topological,huang2024hyperbolic,hashemi2024floquet,tusnin2023nonlinear,hashemi2025reconfigurable,jalali2023topological}, which can enable the design of optimal architectures for maximally relaxed FPM—beyond single-timescale systems~\cite{hu2022photo,clementi2023chip,nitiss2022optically,clementi2024ultrabroadband,zhou2024self}. Another intriguing direction can be extending our approach to higher-order nesting with more than two timescales, for example, offering an efficient pathway to multi-scale photogalvanic effects, multi-harmonic generation~\cite{tonkaev2024even,wang2021high,moore2011continuous}, as well as spatio-temporal FPM~\cite{zhou2025self}. 

\section{Methods}

\subsection{M1: Device fabrication}

The two-timescale photonic devices used in this work were fabricated at a commercial foundry, following the same process as described in Refs.~\cite{flower2024observation,xu2025chip}. A high-resolution optical image of the fabricated two-timescale photonic lattice is shown in Figure~S1. The microring resonators are composed of Si$_3$N$_4$ embedded in a SiO$_2$ cladding. Each ring has a cross-sectional dimension of 1200~nm in width and 800~nm in thickness. The coupling gaps between adjacent microrings, as well as between the microrings and the bus waveguides, are set to 300~nm. 

\subsection{M2: Two-timescale characteristics}

The mode spacing of individual rings $\Omega_{\rm F} = 1/\tau_{\rm F} = \frac{c}{n_{g}L}$, where $c$, $n_{g}$ and $L$ are the speed of light in vacuum, group index and length of the SiN ring, respectively. 
Independently, the edge state spacing in the super-ring lattice $\Omega_{\rm S} = 1/\tau_{\rm S} \simeq \frac{N}{J}$, where $N$ and $J$ are the lattice dimension and inter-ring coupling strength, respectively~\cite{mittal2021topological,xu2025chip}.

\subsection{M3: Experimental setup}

For both linear characterization, quasi-CW pumping, and nonlinear spatial imaging, we followed the procedures described in Refs.~\cite{flower2024observation,xu2025chip}. For power analysis, each harmonic was spectrally filtered before detection. The FH was directed to a high-speed photodetector (50~GHz bandwidth), while the SH and TH were measured using avalanche photodiodes. In all cases, the pump signal was removed prior to detection using a notch filter. Further details on the filtering setup are provided in the SI section S1.

\subsection{M4: FDTD simulation}

Linear-regime 3D simulations of devices have been performed using a commercial FDTD simulation tool~\cite{hughes2021perspective}. For the intensity field profile shown in Figure~\ref{Fig:imaging}a, a representative (with slightly smaller lattice dimension of 5 x 5) was used for computational feasibility. All the other simulation parameters were chosen according to the actual geometry and add-drop port configuration of the experiments.

\section{Acknowledgments}

The authors wish to acknowledge fruitful discussions with Dirk Englund, Curtis Menyuk, Avik Dutt, and Sunil Mittal. M.J.M., L.X., S.K., and M.H. acknowledge Flexcompute for access to simulation resources based on their Tidy3D software. Certain commercial equipment, instruments, or materials (or suppliers, or software, ...) are identified in this paper to foster understanding. Such identification does not imply recommendation or endorsement by the National Institute of Standards and Technology, nor does it imply that the materials or equipment identified are necessarily the best available for the purpose.

\section{Authors contribution}

L.X. and M.J.M. performed the experiments and simulations and analyzed the data. C.J.F.  designed the devices and contributed to the early stages of the setup. M.J.M., L.X., S.S., M.G., A.P., and D.G.S.F. contributed to the construction of the experimental setup. S.S. and M.J.M. performed the FDTD simulations. S.C.O. and G.M. performed the high-Q single-ring measurements. M.J.M. wrote the manuscript. The project was supervised by Y.C., G.M., K.S., and M.H. All authors discussed the results and contributed to the manuscript.

\section{Competing interests}

M.J.M., L.X., G.M., K.S. and M.H. have filed a provisional patent for nested frequency and phase matching for high-yield integrated nonlinear optics. Moreover, M.H. is an inventor on a patent (US Patent 11599006) dated 7 March 2023 that covers the generation of nested frequency combs in a topological source. The authors declare no other competing interests.

\section{Data availability}
All of the data that support the findings of this study are reported in the main text and Supplementary Information. Source data are available from the corresponding authors on reasonable request.

\newpage
\section{Supplementary Materials for\\ Multi-timescale Frequency–Phase Matching for High-Yield Nonlinear Photonics}

\newpage
\setcounter{figure}{0}
\section{S1: Experimental setup}

\begin{figure*}[b]
    \centering
    \includegraphics[width=0.85   \textwidth]{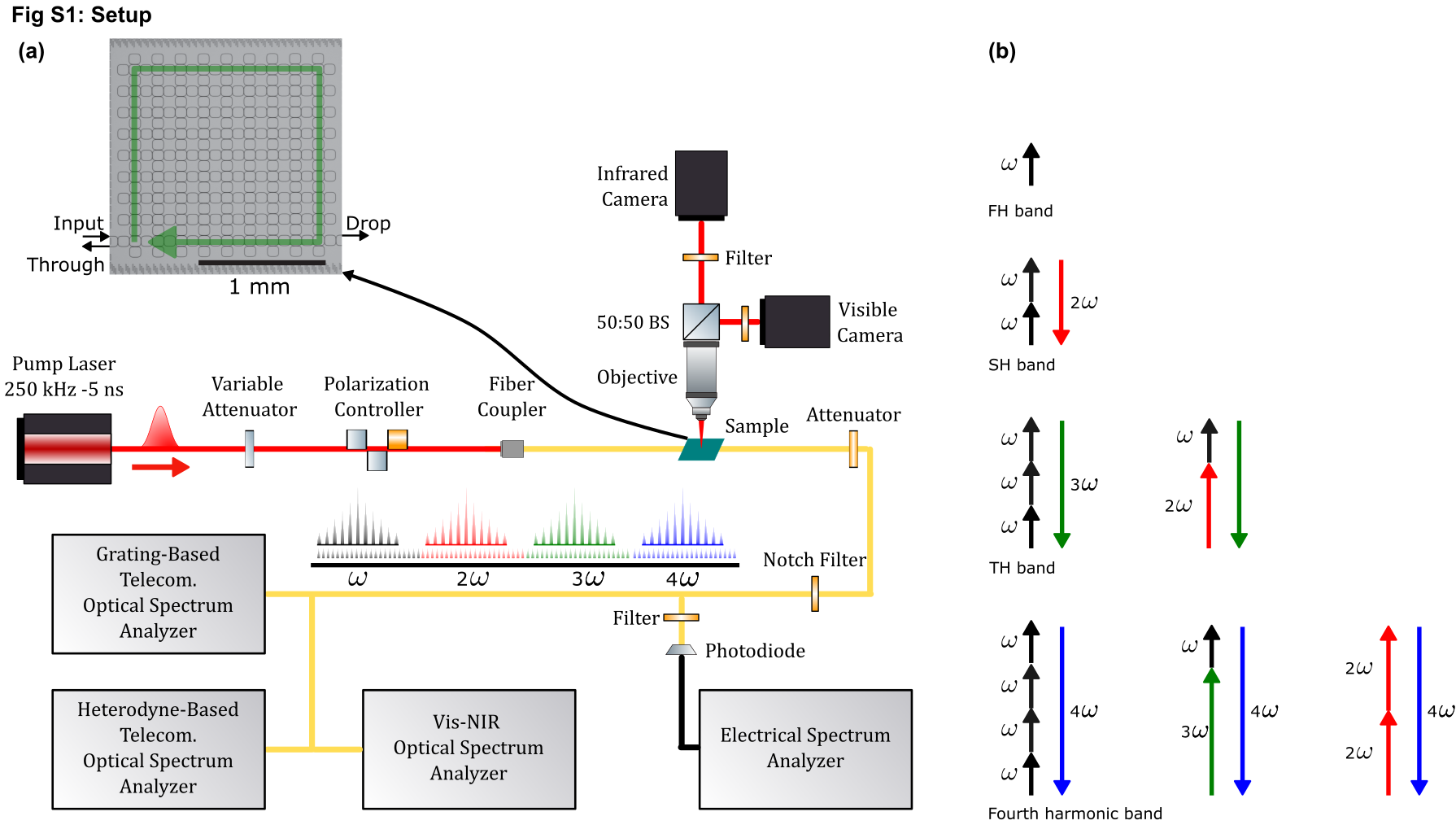}
    \caption{(a) Detailed schematic of the measurement setup. A tunable pulsed laser operating at telecommunication wavelengths passes through a variable attenuator and a polarization controller before being fiber-coupled into the SiN device. The harmonics are collected via fiber, optionally passed through a second variable attenuator and a notch filter for pump suppression, and then routed to the grating-based and heterodyne-based OSAs, and ESA for analysis. Simultaneously, the harmonics are imaged from above using a 10$\times$ objective followed by a 50:50 beam splitter (BS), with one optical path directed to a visible camera and the other reaching an IR-sensitive camera (both are filtered by harmonic-selective filters prior to the camera). A high-quality optical image of the sample is also shown at the top left. (b) The available pure harmonic as well as sum-frequency processes inside the device for the generation of light in each harmonic band. }
    \label{Fig:setup}
\end{figure*}

Figure~\ref{Fig:setup} illustrates the (a) experimental setup used for multi-harmonic generation and detection as well as (b) available nonlinear pathways for the generation of light in each harmonic band. A high-resolution optical image of one of the devices is also shown. A pulsed tunable laser is routed through a free-space system incorporating a variable attenuator and a polarization controller before coupling into a short tapered fiber and edge-coupling into the SiN chip via the input port. Coupling losses are estimated at 2~dB to 3~dB per facet. The combined pump and comb output is collected from the drop port, optionally attenuated or filtered, and analyzed using photo-detectors, broadband (400~nm to 1800~nm) low-resolution (20~pm) and narrow-band (1510~nm to 1630~nm) high-resolution (40~fm) OSAs (grating-based and heterodyne, respectively), and an electrical spectrum analyzer (ESA).

For multi-harmonic spatial imaging, out-of-plane scattered light is collected with a 10$\times$ objective (NA=0.28) and directed through a 50:50 beam splitter to both a visible and an IR-sensitive camera. Filters are routinely used for harmonic-selective imaging, as discussed in the main.

\section{S2: FDTD simulation of the Waveguide Mode Profiles}

\begin{figure*}[h]
    \centering
    \includegraphics[width=0.9\textwidth]{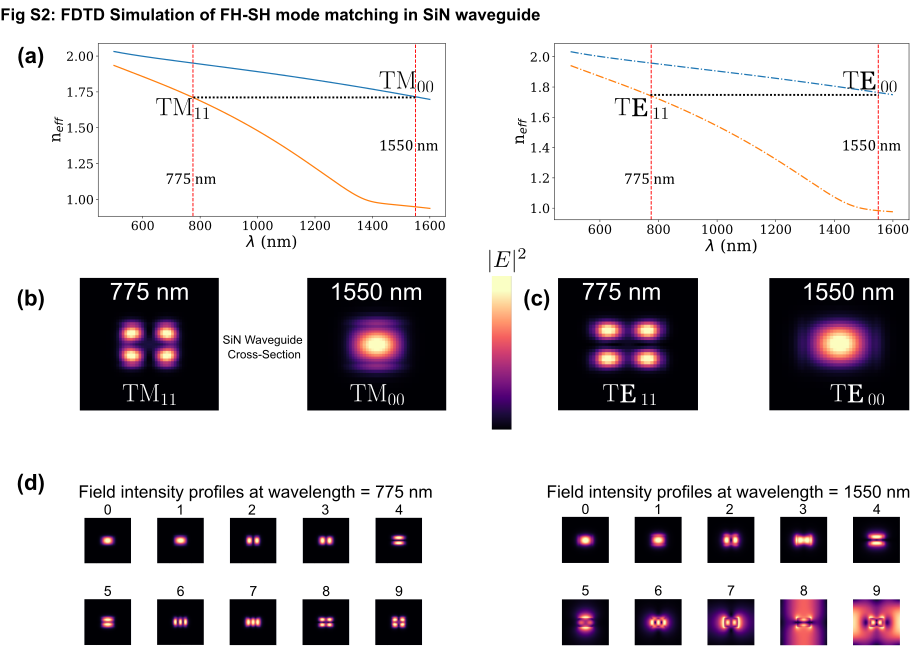}
    \caption{(a-c) FDTD simulation of electric field intensity profiles of the SiN waveguide modes (for our SiN waveguide's cross-sectional dimension of 1200~nm in width and 800~nm in thickness) as a function of wavelength and effective group index. The field profiles are shown at the pump and SH wavelengths. (d) Corresponding higher-order modes of the waveguide. Transverse-electric (TE) and transverse-magnetic (TM) modes are indexed by even and odd numbers, respectively.}
    \label{Fig:FPM}
\end{figure*}

The linear-regime FDTD simulation of field profiles of the SiN waveguide modes as a function of wavelength and effective group index is shown in Figure~\ref{Fig:FPM}. Here, the FPM windows for the fundamental modes at the pump wavelength and mode, and those of the SH are considered. Higher-order modes are also shown for completeness.

\newpage
\section{S3: Tight-binding simulation of linear-regime transmission spectra and field intensity profiles}

\begin{figure*}[h]
    \centering
    \includegraphics[width=0.9\textwidth]{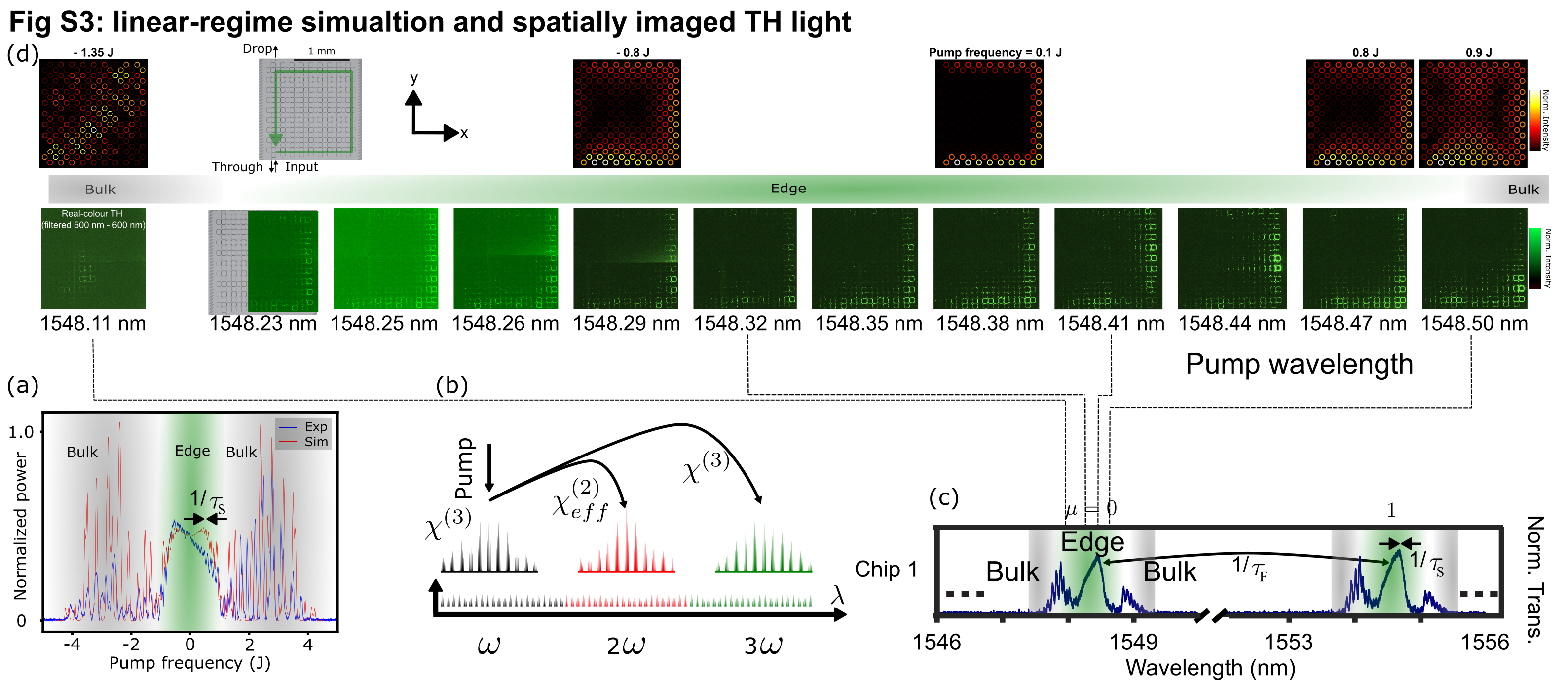}
    \caption{(a) Simulated (red) and measured (blue) linear-regime (weak pump power) transmission spectrum at the drop port of the device for a single longitudinal mode of the lattice. The x-axis denotes the inter-ring coupling strength $J$. b) Schematic of the nonlinear processes for harmonic generation. (c) Linear drop-port transmission spectra of chip 1. (d) A few selected linear-regime simulated spatial intensity profiles and corresponding spatial images of TH light.}
    \label{Fig:sim}
\end{figure*}

Following the linear Hamiltonian description and the tight-binding method discussed in Refs.~\cite{flower2024observation,xu2025chip}, we calculate the single-mode linear simulations of the drop port transmission spectrum, which together with the corresponding experimental data for a representative device (chip 1) are shown in Figure~\ref{Fig:sim}a. Moreover, a few selected spatial intensity profiles of edge and bulk modes, as well as corresponding imaged TH light, are shown. One can see the evolution of confinement of the edge modes by moving the pump wavelength away from the center of the edge band, eventually entering the bulk band with distinct bulk profiles.

\newpage
\section{S4: Electrical spectrum analysis of the frequency comb}

\begin{figure*}[h]
    \centering
    \includegraphics[width=0.7\textwidth]{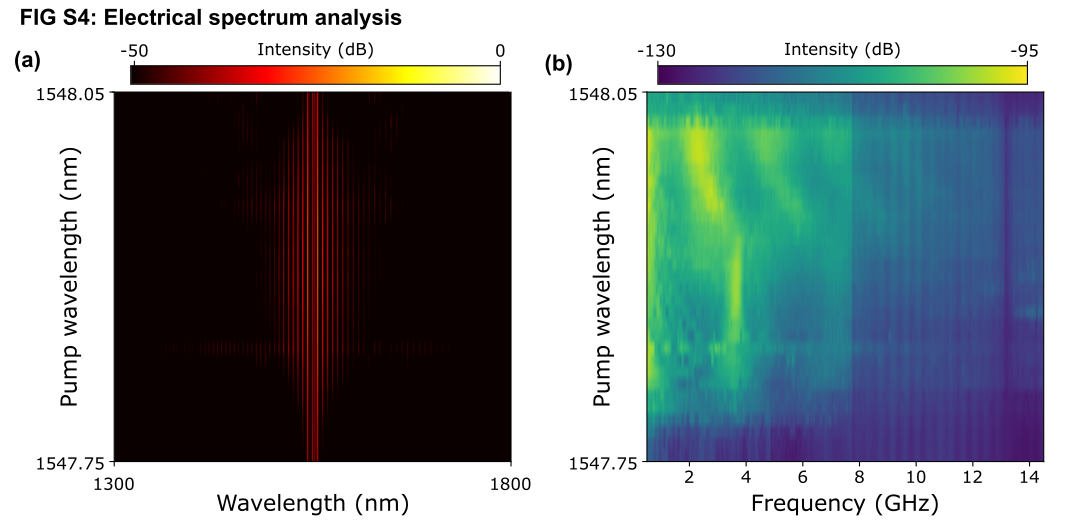}
    \caption{(a) Optical spectrum and (b) corresponding electrical noise analysis of the combs in Chip 3 as a function of pump wavelength within the edge using 200~mW of pump power. In (b), the beats correspond to the slow-timescale repetition rate of the combs. The pump is notched out before sending the spectrum to the OSA and ESA. In (b), the RF output of the photodiode was sent to the ESA.}
    \label{Fig:ESA}
\end{figure*}

To investigate the noise properties of the combs, we use a fast (43.5~GHz) electrical spectrum analyzer (ESA) and analyze the RF output of the 50~GHz photodiode in our setup. The measured OSA and ESA spectra for combs generated in chip 3 are shown in Figure~\ref{Fig:ESA} as a function of pump wavelength within the edge band. The pump was notched out in all of these measurements. We observe similar noise properties and slow-time-scale comb beats as the ones recently reported in Ref.~\cite{xu2025chip}.

\newpage
\section{S5: Optical power analysis of the harmonics}

\begin{figure*}[h]
    \centering
    \includegraphics[width=0.9\textwidth]{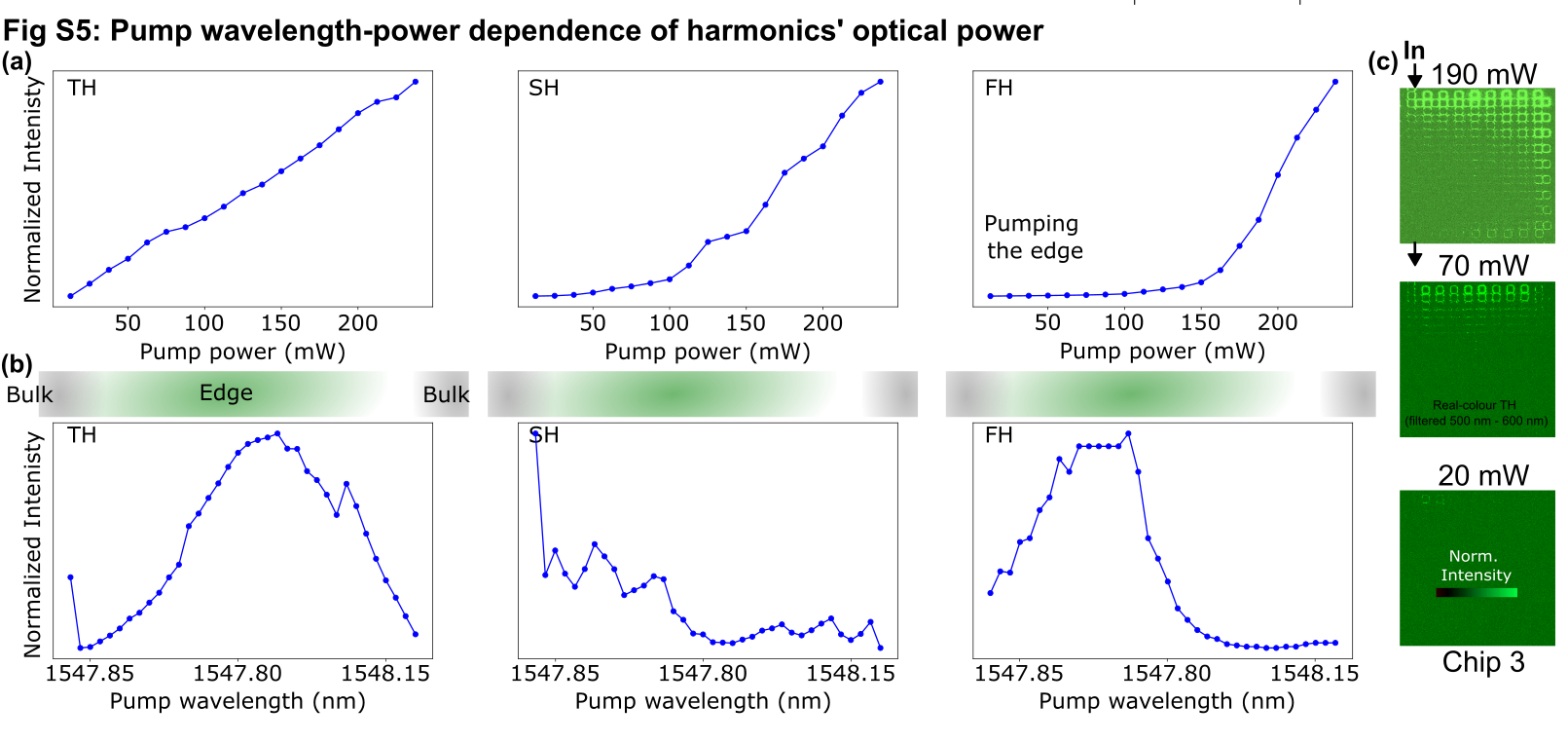}
    \caption{Spectrally-filtered independent optical power analysis of the nested fundamental, second, and third harmonics (FH, SH, and TH) as a function of pump \textbf{(a)} power (generated when pumping the center of the edge band) and \textbf{(b)} wavelength (with approximately 185 mW of average pump power). The bulk and edge regions of the spectrum are highlighted with gray and green, respectively. In each panel, all the data are normalized to the maximum value. \textbf{(c)} Pump-power dependence of the spatially-imaged TH light for a typical edge mode. The input and drop ports are marked with arrows on the top panel. At 190 mW pump power, scattered light can be seen all the way from the input to the drop port.}
    \label{Fig:power}
\end{figure*}

The pump wavelength and pump power dependence of the nested harmonics optical power are summarized in Figure~\ref{Fig:power}a-b. The top (bottom) row shows the measured harmonics powers as a function of pump power (pump wavelength). The same filtering configuration as the one in the spatial imaging (Figure 2 of the main text) is used. For the results in Figure~\ref{Fig:power}a, the pump wavelengths was fixed at the center of the edge band. In Figure~\ref{Fig:power}b, the pump power was fixed at 190 mW for wavelength sweep over the same wavelength range of the nonlinear measurements. We observe that the edge FH and TH have higher power than their bulk counterparts. However, the SH power shows a more complex behavior. This behavior is likely due to the distinction between the all-optically induced $\chi^{(2)}_{\mathrm{eff}}$ process driving SHG and the intrinsic $\chi^{(3)}$ nonlinearity underlying FWM in the FH band and TH generation in SiN. The pump-power dependence of the spatially imaged TH light is shown in Figure~\ref{Fig:power}c. We note that the pump was filtered out in all of these measurements. 

\newpage
\section{S6: Multi-harmonic imaging}

\begin{figure*}[h]
    \centering
    \includegraphics[width=0.7\textwidth]{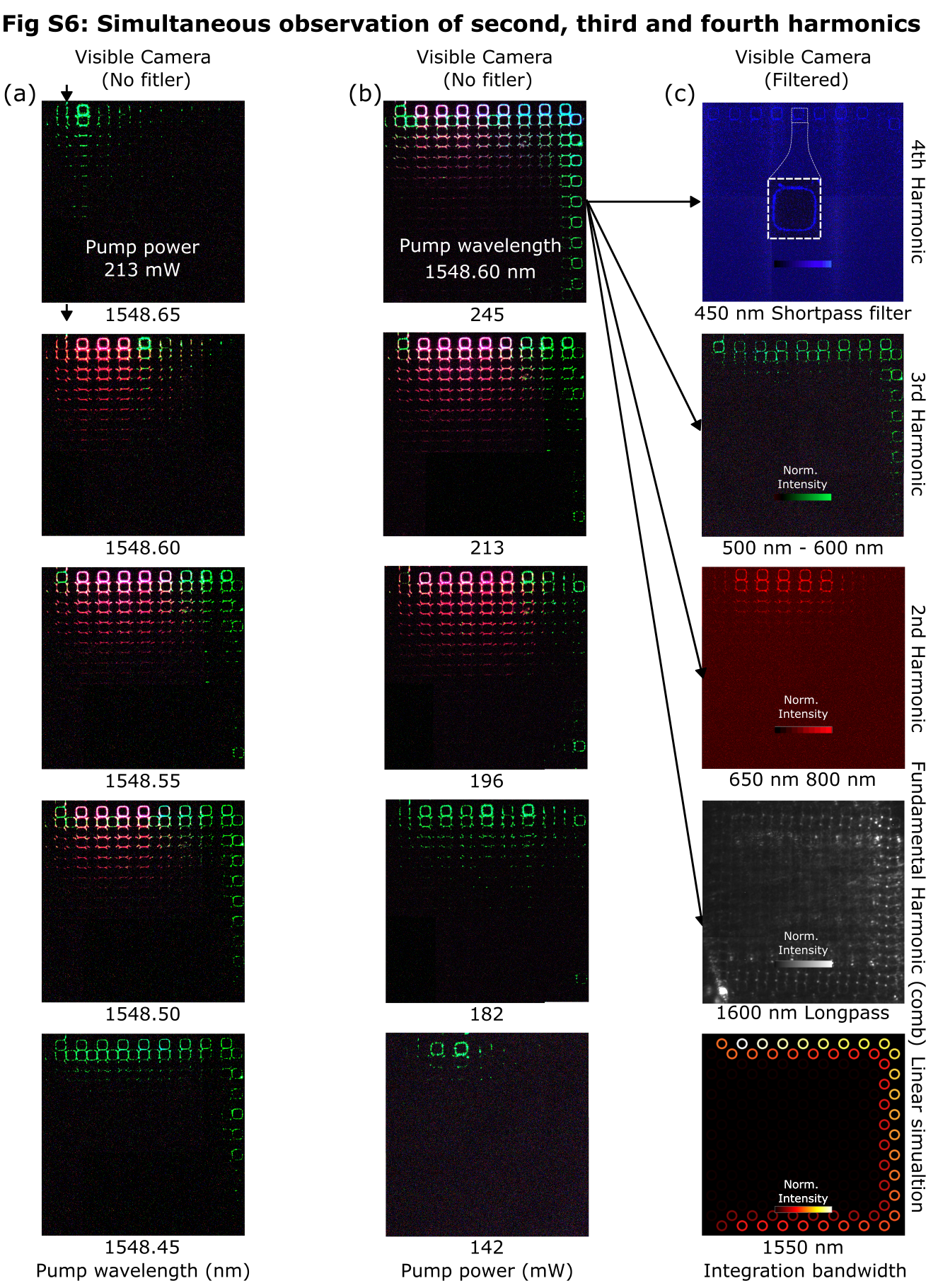}
    \caption{Unfiltered real-color (a) pump-wavelength and (b) pump-power dependence of spatial imaging of the harmonics. (c) Simulated linear intensity profile of a typical edge mode, followed by real-color spectrally filtered FH, SH, TH, and the fourth harmonics, respectively.}
    \label{Fig:multi-hhg}
\end{figure*}

An extended version of the spatial imaging of the multi-harmonics (presented in the main, Figure 2) is shown in Figure~\ref{Fig:multi-hhg}a-c. Specifically, here we include the pump-wavelength-dependence of the spatial images for a fixed pump power of 213~mW, shown in Figure~\ref{Fig:multi-hhg}a.

\newpage
\section{S7: Pump-wavelength-dependent spatial imaging of the FH light}

\begin{figure*}[h]
    \centering
    \includegraphics[width=1\textwidth]{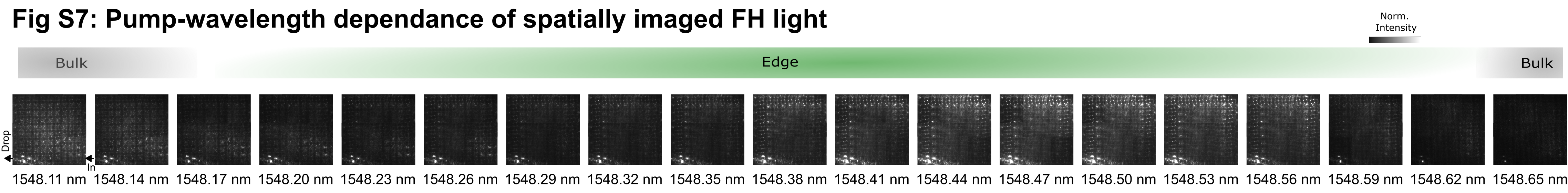}
    \caption{(a) Pump wavelength dependence of the spatial imaging of FH light in chip 1. The bulk and edge parts of the spectrum are marked with gray and green, respectively. For all cases, a fixed pump power of 200~mW was used, and all images were taken using a 1600 nm longpass filter.}
    \label{Fig:FHimage}
\end{figure*}

To investigate the robustness of FH generation as a function of pump wavelength within the edge band, we image the FH light in chip 1 within its entire edge band as well as part of its bulk band (similar to the TH imaging presented in the main Figure 4). The results are shown in Figure~\ref{Fig:FHimage}. We observe that within the entire edge band, the FH light is confined to the edges of the lattice and circulates around the sharp corners. In contrast, the FH light generated when pumping the bulk originates from the interior of the lattice and lacks the edge characteristics. These observations are consistent with both spatial imaging of the TH light as well as the simulated linear intensity profiles discussed in the section S3.

\newpage
\section{S8: Pseudo-spin and bulk comparison}

\begin{figure*}[h]
    \centering
    \includegraphics[width=0.8\textwidth]{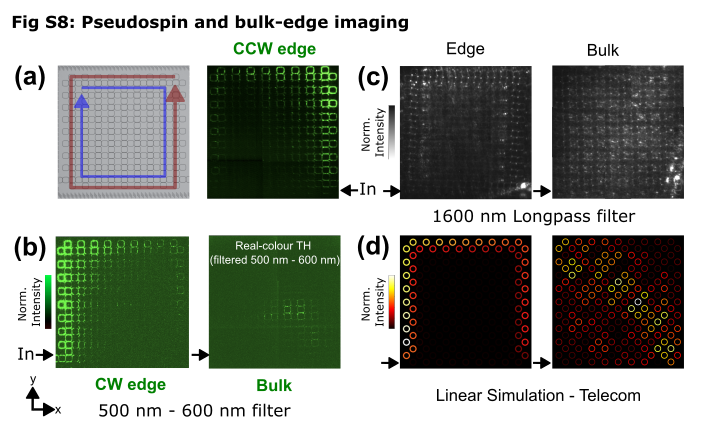}
    \caption{(a) Left panel: optical image of the device, where CCW and CW edge channels are marked by red and blue, respectively. Right panel: spatial imaging of filtered TH light for CCW and (b) CW edge excitation, respectively. The right panel of (b) shows TH light generated when pumping the bulk band. (c) Corresponding filtered (using a 1600 longpass filter) FH light for the cases in (b). (d) Corresponding linear intensity profile for the edge and bulk modes.}
    \label{Fig:ccw}
\end{figure*}

To explore the CCW and CW (pseudo-spin) degrees of freedom as well as the contrast between the bulk and edge excitation scenarios, we perform FH and TH spatial imaging for each case. As presented in Figure~\ref{Fig:ccw}a-b, both pseudo-spin channels show clear edge-confined propagation of the TH light. Figure~\ref{Fig:ccw}c shows the corresponding FH light. Moreover, Figure~\ref{Fig:ccw}b-c contrasts the bulk versus edge excitation cases for both TH and FH. By comparing these observations with the corresponding simulated linear intensity profiles shown in Figure~\ref{Fig:ccw}d, we can see that the linear-regime edge and bulk characteristics of the modes are retained at the present harmonic processes.

\newpage
\section{S9: Analyzing the nestedness of the frequency combs with heterodyne-based OSA}

\begin{figure*}[h]
    \centering
    \includegraphics[width=0.9\textwidth]{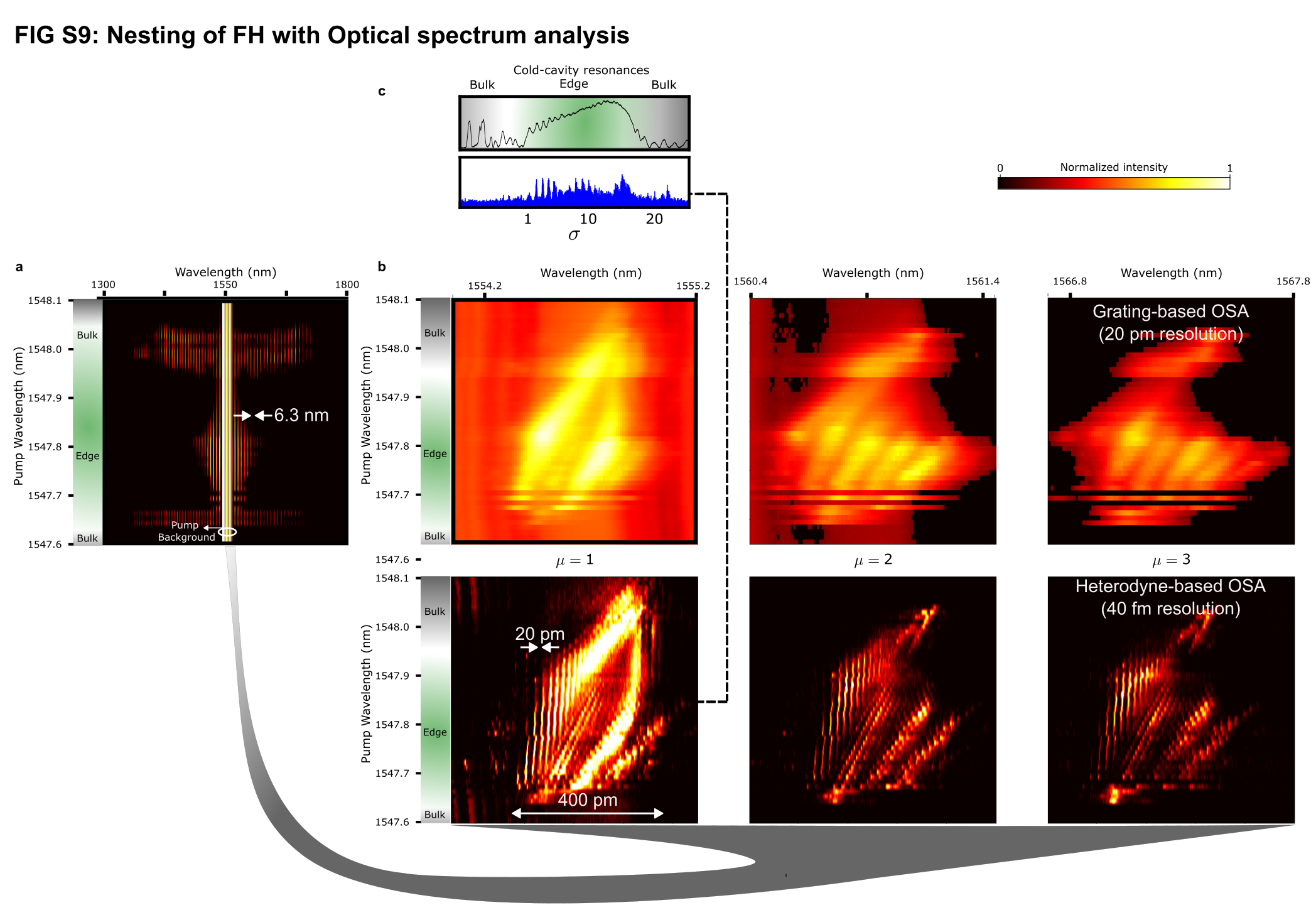}
    \caption{Broadband grating-based (a) and selected narrow-band (b) heterodyne-based OSA analysis of the comb generation in Chip 1. The spectral resolution of the grating-based (heterodyne-based) OSA is 20 pm (40 fm) in (b), respectively. The nested structure of the comb teeth can clearly be seen in the bottom row of (b). (c) A snapshot of a typical nested comb teeth and the corresponding cold-cavity linear drop-port transmission for comparison. For each fast-timescale longitudinal mode $\mu$, approximately 20 slow-timescale modes $\sigma$ can be seen, comprising the predicted 400 pm edge bandwidth.}
    \label{Fig:apex}
\end{figure*}

Since our grating-based based OSAs, which were used for the data presented in the main text, have a limited spectral resolution of 20 pm, we use a narrow-band (1520~nm to 1630~nm) heterodyne-based OSA with 40 fm resolution to directly confirm the nestedness of the harmonics (in this case, the FH). The results are summarized in Figure~\ref{Fig:apex}. Panel a shows the pump-wavelength dependence of the combs in Chip 1 generated with a fixed pump power of 200~mW and measured with a broadband grating-based OSA. The fast-timescale FSR of $\approx$ 6.3~nm is marked. We focus on three representative longitudinal modes $\mu =1$ to $\mu =3$, as shown in the top row of Figure~\ref{Fig:apex}b. The bottom row shows the corresponding heterodyne-based measured spectra, clearly showing $\approx$ 20 slow-timescale longitude modes $\sigma$ within a 400~pm edge band at each $\mu$ and spaced with a 20~pm FSR as expected. Figure~\ref{Fig:apex}c shows a typical snapshot of a nested comb tooth (bottom panel) and the corresponding cold-cavity linear drop port spectrum for comparison.

\newpage
\section{S10: Harmonic generation in low-Q single-timescale single-ring resonators}

\begin{figure*}[b]
    \centering
    \includegraphics[width=0.6\textwidth]{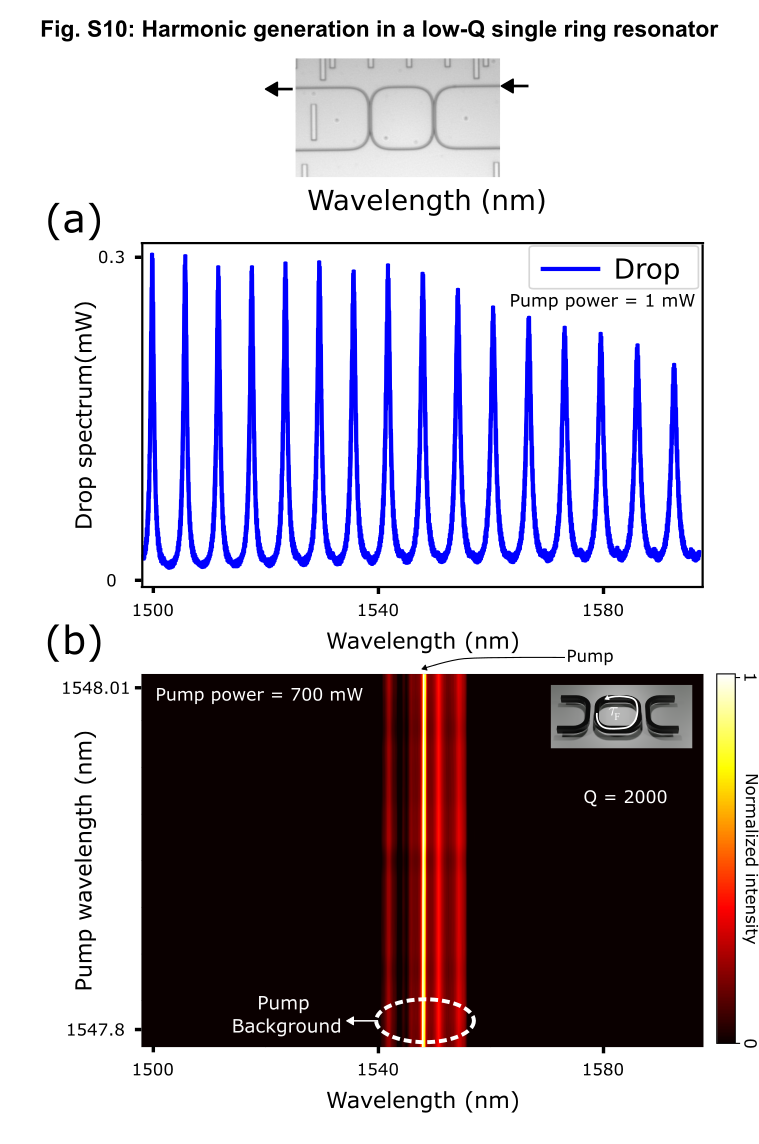}
    \caption{(a) Linear-regime drop-port transmission of a low-Q (2000) single-ring resonator. The FSR is 750 GHz (6.3 nm), same as the fast timescale of the lattice counterpart. A high-quality image of the device is also shown. (b) Absence of comb generation in the single ring pumped with the maximum power of the pump (off-chip 700 mW).}
    \label{Fig:ADF10}
\end{figure*}

To directly compare our relaxed FPM approach to single-timescale counterparts, we study the single-ring version of our lattice. Specifically, on one of the chips studied in the main study, we investigate harmonic generation on a single-ring SiN device designed with the same add-drop filter configuration. We note that for optimal comparability, we used exactly the same pump and ADF ring geometry (the bending type, ring length, and 300 nm gaps for the bus waveguides, etc). The linear drop-port characterization of the device (using a tunable CW laser with 1 mW of power) is shown in Figure~\ref{Fig:ADF10}a, showing modes with approximately 2000 Q-factor at telecommunication wavelength. 

First, we investigate FH generation in the ring. We use the maximum off-chip power of our laser (700 mW) and sweep the pump wavelength around the center of the mode at 1547 nm. As shown in Figure~\ref{Fig:ADF10}b, despite such high pump power, we do not observe any comb formation due to the low Q factor and the corresponding high OPO threshold.

\begin{figure*}[h]
    \centering
    \includegraphics[width=0.99\textwidth]{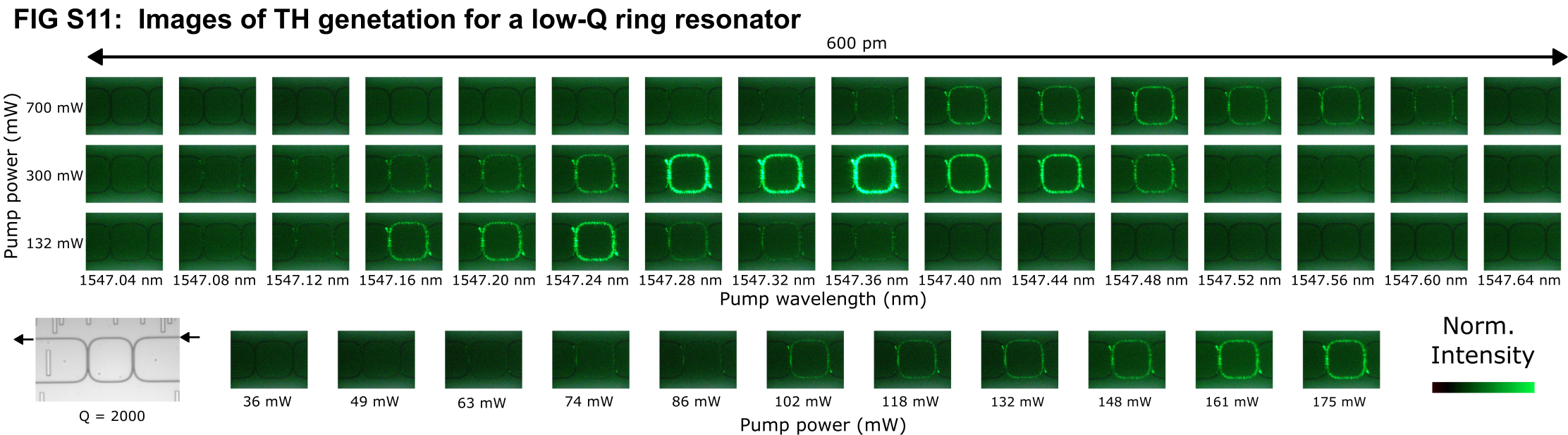}
    \caption{TH generation in a low-Q (2000) single-ring counterpart of our lattice. Bottom row shows the pump power dependence of the spatial imaging of the filtered TH light from the single-ring when pumping the center of the ring mode near 1547 nm. The top three rows show the pump wavelength dependence of the TH light for three pump powers of 132 mW, 300 mW, and 700 mW, respectively. Thermal shift of the resonance can be clearly seen, which at 700 mW pump power, leads to deviation from optimal FPM and reduction of the TH light.}
    \label{Fig:ADF10TH}
\end{figure*}

Next, we investigate THG in the low-Q single-ring shown in Figure~\ref{Fig:ADF10}. The results for the spatial imaging of the filtered TH light are summarized in Figure~\ref{Fig:ADF10TH}. First, by pumping the center of the mode near 1547 nm at different pump powers (up to 175 mW as shown in the bottom row), we observe the generation of TH light proportional to the pump power. 

Next, for the fixed pump powers of 132 mW, 300 mW, and 700 mW, we sweep the pump wavelength over a range (1547.04 nm to 1547.60 nm) covering the mode resonance. We observe that while the central resonance of the TH mode shifted to longer wavelengths as a function of increased pump power (due to thermal shift), the TH light maximum intensity (after increasing when pump power is increased from 132 mW to 300 mW) drops dramatically for pump power of 700 mW. This serves as a clear indication of the deviation of the FPM condition between the pump mode and the TH mode in this single-timescale counterpart of our system.

\newpage
\section{S11: Low-yield harmonic generation in High-Q single rings}

\begin{figure*}[h]
    \centering
    \includegraphics[width=0.7\textwidth]{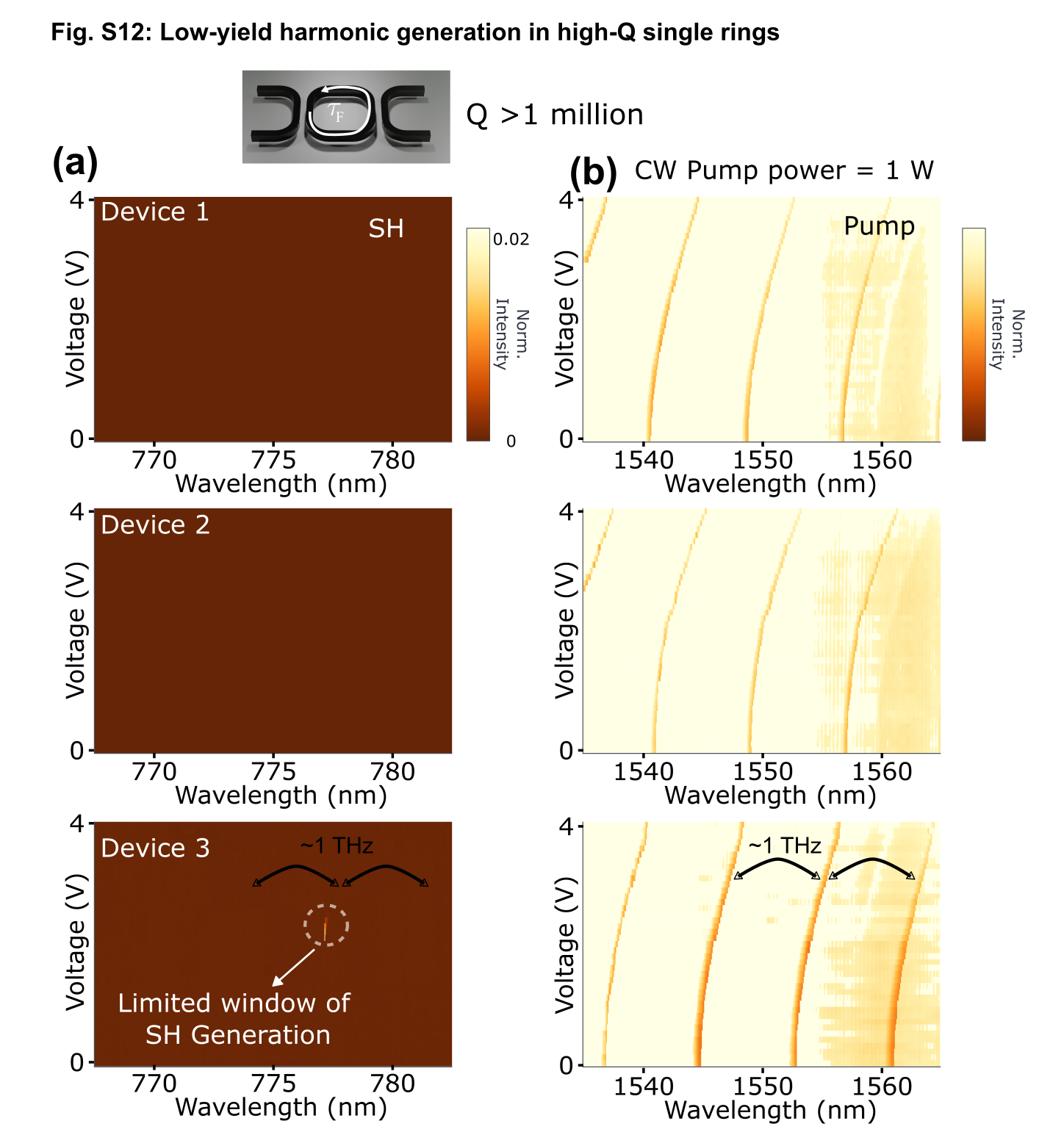}
    \caption{SH generation in high-Q (exceeding one million) SiN single rings. (a) Measured SH signal in three devices and (b) the corresponding measured transmission spectrum at the pump wavelength using off-chip pump power of 1 W. The y-axis indicates the voltage of the thermal heaters used to tune the ring resonances. SHG is only observed for one of the devices, for a very small FPM window, and only for one of the modes.}
    \label{Fig:highq}
\end{figure*}

Satisfying the FPM conditions for harmonic generation is typically even more difficult in higher Q-factor ring resonators, due to the much narrower linewidth of the modes. To investigate this, we study SHG in three SiN single-ring resonators with Q factors exceeding one million and a 1 THz FSR (comparable to our lattice's fast time scale of 750 GHz). We use heaters to tune the ring modes and explore potential windows that satisfy SHG's FPM conditions. The results, (using a 1 W off-chip CW high-power pump) are summarized in Figure~\ref{Fig:highq}. We observe three clear signatures of stringent FPM conditions. First, we observe a very low functional device yield for the generation of SH, noting that only device 3 shows SHG. Second, SHG in device 3 is only observed for a very small window, indicating the strict constraints required for such nonlinear optical processes. Third, only one of the modes exhibits SHG. All these observations are in sharp contrast to our nested FPM, which passively enables high-yield, relaxed, and broad multi-mode harmonic generation.

\newpage
\bibliography{Main.bib}

\end{document}